\documentstyle[11pt,epsfig]{article}

\newcommand{\ra}{\rightarrow}

\newcommand{\s}{\\ \vspace*{-3mm} }
\newcommand{\nn}{\noindent}

\newcommand{\beq}{\begin{eqnarray}}
\newcommand{\eeq}{\end{eqnarray}}

\def\ti              {\tilde}

\def\st              {{\ti t}}

\def\sb              {{\ti b}}

\def\stau            {\ti \tau}

\newcommand\tht{{\theta_\st}}

\def\stau            {\ti \tau}

\begin{document}
\hfill IFT/99-05\\
\null
\hfill hep-ph/9904260\\
\vskip .8cm
\begin{center}
{\Large \bf Supersymmetry Searches at $e^+e^-$ Linear
Colliders\footnote{Presented at the Cracow Epiphany Conference, Cracow,
January 1999} 
\\[.5em]
}
\vskip 2.5em

{\large  Jan Kalinowski}
\centerline{ Instytut Fizyki Teoretycznej UW, Ho\.za 69, 00681 Warsaw, Poland}
\end{center}

\vspace*{1cm}
\begin{center}
{\bf Abstract}
\end{center}
The physics potential of discovering and exploring supersymmetry at
future $e^+e^-$ linear colliders is reviewed.  Such colliders are
planned to start to operate at a center--of--mass energy of 500~GeV to
800~GeV, with a final energy of about 2~TeV expected.  They are ideal
facilities for the discovery of supersymmetric particles.  High
precision measurements of their properties and interactions will help
to uncover the mechanism of supersymmetry breaking and will allow for
tests of grand unification scenarios.

\null
\clearpage

\newpage
  
\section{Introduction}
The Standard Model is exceedingly successful in describing leptons,
quarks and their interactions.  Nevertheless, the Standard Model (SM)
is not considered as the ultimate theory since neither the fundamental
parameters, masses and couplings, nor the symmetry pattern are
predicted. These elements are merely built into the model.  Likewise,
the spontaneous electroweak symmetry breaking is simply parametrized
by a single Higgs doublet field.

Even though many aspects of the Standard Model are experimentally
supported to a very high accuracy \cite{Hollik}, the embedding of the
model into a more general framework is to be expected.  The argument
is closely related to the mechanism of the electroweak symmetry
breaking.  If the Higgs boson is light, the Standard Model can
naturally be embedded in a grand unified theory.  The large energy gap
between the low electroweak scale and the high grand unification scale
can be stabilized by supersymmetry. Supersymmetry \cite{susy} actually
provides the link between the experimentally explored interactions at
electroweak energy scales and physics at scales close to the Planck
scale where the gravity is important.  If the Higgs boson is very
heavy, or if no fundamental Higgs boson exists, new strong
interactions between the massive electroweak gauge bosons are
predicted by unitarity at the TeV scale. With possibly many more new
layers of matter before the Planck scale is reached, no direct link
between electroweak and Planck scales in such a scenario is expected
at present.  In either case, the next generation of accelerators which
will operate in the TeV energy range, can uncover the structure of
physics beyond the Standard Model.

Despite the lack of direct experimental evidence\footnote{The status
of low-energy supersummetry is discussed by S. Pokorski \cite{SP}.}
for supersymmetry (SUSY), the concept of symmetry between bosons and
fermions has so many attractive features that the supersymmetric
extension of the Standard Model is widely considered as a most natural
scenario. SUSY ensures the cancellation of quadratically divergent
quantum corrections from scalar and fermion loops and thus the Higgs
boson mass can be kept in the desired range of order $10^2$ GeV, which
is preferred by precision tests of the SM. The prediction of the
renormalized electroweak mixing angle $\sin^2 \theta_W$, based on the
spectrum of the Minimal Supersymmetric Standard Model (MSSM), is in
striking agreement with the measured value.  Last but not least,
supersymmetry provides the opportunity to generate the electroweak
symmetry breaking radiatively.

In the next section the silent features of supersymmetric models are
briefly summarized. We will stress the importance of determining
experimentally all SUSY parameters in a model independent way.  For
this purpose the $e^+e^-$ linear colliders \cite{LC} are indispensable
tools.  It is illustrated in the next chapter where some of the
recently developed strategies to ``measure'' SUSY parameters in the
gaugino and sfermion sectors are discussed. For a discussion of the
SUSY Higgs sector we refer to \cite{PMZ}.

\section{Low-energy MSSM} 

Supersymmetry predicts the quarks and leptons to have scalar partners,
called squarks and sleptons, the gauge bosons to have fermionic
partners, called gauginos. In the MSSM \cite{mssm} two Higgs doublets
with opposite hypercharges, and with their superpartners -- higgsinos
-- are required to give masses to the up and down type fermions and to
ensure anomaly cancellation. Thus the particle content of the MSSM is
given by

\vspace{-0.5cm}
\begin{table}[h]
\begin{center}
\begin{tabular}{|ccccc|cccc|cc|}
\hline \rule{0cm}{4mm}
$(l_a,\nu_a)_L$ & $l^c_{aL}$ &$(u_a,d_a)_L$ &$u^c_{aL}$&$d^c_{aL}$ 
&$\gamma$& $W^\pm$ &$Z^0$ &$g_{i}$&$H_1$& $H_2$
\\ \hline \rule{0cm}{4mm}
$({\tilde l}_a,{\tilde\nu}_{a})_L$ &${\tilde l}^c_{aL}$ &$
({\tilde u_a},{\tilde d_a})_L$ & ${\tilde u}^c_{aL}$&${\tilde d}^c_{aL}$  
&${\tilde\gamma}$& ${\tilde W}^\pm$ &${\tilde Z}^0$ &
${\tilde g}_{i}$ &${\tilde H}_1$& ${\tilde H}_2$ \\
\hline
\end{tabular}
\end{center}
\end{table}

\vspace{-0.5cm} 
\noindent where the first row lists the (left-handed) fermion fields
of one generation ($a=$1-3), the gauge fields (for gluons $i=$1-8) and
two Higgs doublets, and the second row -- their superpartners.  The
higgsinos and electroweak gauginos mix; the mass eigenstates are
called charginos and neutralinos for electrically charged and neutral
states, respectively.  The MSSM is defined by the superpotential
\begin{equation}
  W=Y_{ab}^e \hat{L}_a\hat{H}_1 \hat{E}^c_b + Y_{ab}^d \hat{Q}_a 
  \hat{H}_1\hat{D}^c_b
  +Y_{ab}^u\hat{Q}_a\hat{H}_2\hat{U}^c_b 
   -  \mu \hat{H}_1 \hat{H}_2 \label{Rcons}
\end{equation}
where standard notation is used for the superfields of left-handed
doublets of (s)leptons ($\hat{L}_a$) and (s)quarks ($\hat{Q}_a$), the
right-handed singlets of charged (s)leptons ($\hat{E}_a$), up-
($\hat{U}_a$) and down-type (s)quarks ($\hat{D}_a$), and for the Higgs
doublet superfields which couple to the down ($\hat{H}_1$) and up
quarks ($\hat{H}_2$); the indices $a,b$ denote the generations and a
summation is understood, $Y^f_{ab}$ are Yukawa couplings and $\mu$ is
the Higgs mixing mass parameter. The $W$ respects a discrete
multiplicative symmetry under $R$-parity, defined as $
R_p=(-1)^{3B+L+2S} $, where $B$, $L$ and $S$ denote the baryon and
lepton number, and the spin of the particle. The $R_p$ conservation
implies that the lightest supersymmetric particle (LSP -- preferably
the lightest neutralino) is stable and superpartners can be produced
only in pairs in collisions and decays of particles.

If realized in Nature, supersymmetry must be broken at low energy
since no superpartners of ordinary particles have been observed so
far.  It is technically achieved \cite{gri} by introducing the
soft--supersymmetry breaking \\ 
(i) gaugino mass terms for bino
$\tilde{B}$, wino $\tilde{W}^j$ $[j=$1--3] and gluino $\tilde{g}^a$
$[i=$1--8] \beq {\textstyle \frac{1}{2}} M_1
\,\overline{\tilde{B}}\,\tilde{B} \ + \ {\textstyle \frac{1}{2}} M_2
\,\overline{\tilde{W}}^i \,\tilde{W}^i \ + \ {\textstyle \frac{1}{2}}
M_3 \,\overline{\tilde{g}}^i \,\tilde{g}^i \ , \eeq (ii) trilinear
couplings (generation indices are understood) \beq A^u H_2 \tilde{Q}
\tilde{u}^c + A^d H_1 \tilde{Q} \tilde{d}^c + A^l H_1 \tilde{L}
\tilde{l}^c - \mu B H_1 H_2 \eeq (iii) and squark and slepton mass
terms \beq m_{\tilde{Q}}^2 [\tilde{u}^*_L\tilde{u}_L
+\tilde{d}^*_L\tilde{d}_L] + m_{\tilde{u}}^2 \tilde{u}^*_R \tilde{u}_R
+ m_{\tilde{d}}^2 \tilde{d}^*_R \tilde{d}_R+
 \ \cdots
\eeq
where the ellipses stand for the soft mass terms for sleptons 
and Higgs bosons. \s

The more than doubling the spectrum of states in the MSSM together
with the necessity of including the SUSY breaking terms gives rise to
a large number of parameters. Even with the $R$-parity conserving and
CP-invariant SUSY sector, which we will assume in what follows, in
total more than 100 new parameters are introduced!  This number of
parameters can be reduced by additional physical assumptions.  The
most radical reduction is achieved in the so called mSUGRA, by
embedding the low--energy supersymmetric theory into a grand unified
(SUSY-GUT) framework by requiring at the GUT scale $M_G$: \s

\nn $(i)$ the unification of the U(1), SU(2) and SU(3) coupling
constants 
\begin{equation} \alpha_3 (M_{\rm G})
= \alpha_2 (M_{\rm G}) = \alpha_1 (M_{\rm G}) =\alpha_G, 
\end{equation} 
$(ii)$ a common gaugino mass $m_{1/2}$.  The gaugino masses $M_i$  at the
electroweak scale are then related through renormalization group
equations (RGEs) to the gauge couplings
\beq M_i =
\frac{\alpha_i(M_Z)}{\alpha_G} m_{1/2} ,
\eeq 
$(iii)$ a universal trilinear coupling $A_G$ 
\beq A_G = A^u (M_{\rm G}) = A^d (M_{\rm G}) =
A^l (M_{\rm G}) ,
\eeq 
$(iv)$ a universal scalar mass $m_0$ 
\beq
m_0&=&m_{\tilde{Q}}=m_{\tilde{u}}=m_{\tilde{d}}= \cdots,
\eeq 
$(v)$ radiative breaking of the electroweak symmetry.\s

\noindent 
The last requirement allows to solve for $B$ and $\mu$ (to within a
sign) once the values of the GUT parameters $m_{1/2}$, $m_0$, $A_G$ as
well as the ratio of the vacuum expectation values of the fields
$H_2^0$ and $H_1^0$, $\tan\beta=v_2/v_1$, are fixed.  As a result, the
mSUGRA is fully specified by $m_{1/2}$, $m_0$, $A_G$, $\tan\beta$ and
sign($\mu$) -- the couplings, masses and mixings at the electroweak
scale are determined by the RGEs \cite{msugra}.

From the experimental point of view, however, 
all low-energy parameters should be 
measured independently of any theoretical assumptions. Therefore 
the experimental program to search for and explore SUSY at present and
future colliders should include the following points: 
\begin{itemize}
\item[(a)] discover supersymmetric particles and measure their quantum
   numbers to prove that they are {\it the} superpartners of standard
   particles,
\item[(b)]
   determine the low-energy Lagrangian parameters,
\item[(c)] verify the relations among them in order to distinguish
   between various SUSY models.
\end{itemize}
If SUSY is at work it will be a matter of days for the LC to discover
the kinematically accessible supersymmetric particles.  Once they are
discovered, the priority will be to measure the low-energy SUSY
parameters independently of theoretical prejudices and then check
whether the correlations among parameters, if any, support a given
theoretical framework, like SUSY-GUT relations.  A clear strategy is
needed to deal with so many a prori arbitrary parameters.  One should
realize, that the low-energy parameters are of two distinct
categories. The first one includes all the gauge and Yukawa couplings
and the higgsino mass parameter $\mu$. They are related by exact
supersymmetry which is crucial for the cancellation of quadratic
divergencies.  For example, at tree-level the $qqZ$,
$\tilde{q}\tilde{q}Z$ gauge and $q\tilde{q}\tilde{Z}$ Yukawa couplings
have to be equal.  The relations among these parameters 
(with calculable radiative corrections) have to be
confirmed experimentally; if not -- the supersymmetry is excluded.
The second category encompasses all soft supersymmetry breaking
parameters: Higgs, gaugino and sfermion masses and mixings, and
trilinear couplings. They are soft in the sense that they do not
reintroduce dangerous quadratic divergencies. Each one should be
measured independently by experiment to shed  light
on the mechanism of supersymmetry breaking.
  
Particularly in this respect (points (b) and (c) above) the $e^+e^-$
linear colliders are invaluable. An intense activity during last
decade in Europe, the USA and Japan on physics at a linear $e^+e^-$
collider \cite{lcwork} 
has convincingly demonstrated the advantages and benefits of
such a machine and its complementarity to the Large Hadron Collider
(LHC).  Many studies have shown that the LHC can cover a mass range
for SUSY particles up to $\sim$ 2 TeV, in particular for squarks and
gluinos \cite{SusyLHC}.  The problem however is that many different
sparticles will be accessed at once with the heavier ones cascading
into the lighter which will in turn cascade further leading to a
complicated picture.  Simulations for the extraction of parameters
have been attempted for the LHC \cite{SusyLHC} and demonstrated that
some of them can be extracted with a good precision. However it must
be stressed that these checks were done with the assumption of an
underlying model, like mSUGRA and it has not been demonstrated so far
that the same can be achieved in a model independent way.

From the practical point of view it is very important that the energy
of the $e^+e^-$ machine can be optimized so that only very few
thresholds are crossed at a time. Another important feature is the
availability of beam polarization as well as a possibility of running
in $e^-e^-$ \cite{emem} or in $e\gamma$ and $\gamma\gamma$
\cite{gamma} modes. Making judicious choices of these features, the
confusing mixing of many final states, unavoidable at the LHC, with
the cascade decays might be avoided and analyses restricted to a
specific subset of processes performed.  The measurements that can be
performed in the Higgs sector are discussed in the talk by P. Zerwas
\cite{PMZ}.  Here I will discuss some methods of extracting SUSY
parameters from the gaugino (chargino/neutralino) and sfermion
sectors. In contrast to many earlier analyses \cite{other}, we will
not elaborate on global fits 
but rather we will discuss attempts at
``measuring'' the fundamental parameters.  Such attempts generically
involve two steps:
\begin{enumerate}
\item[$A$:] from the observed quantities: cross sections, asymmetries 
    etc. \\ $\Longrightarrow$ determine the physical parameters:
    the masses, mixings and couplings of sparticles
\item[$B$:] from the physical parameters \\ $\Longrightarrow$ extract
    the Lagrangian parameters: $M_i$, $\mu$, $\tan\beta$, $A^u$, 
$m_{\tilde{Q}}$ etc.
\end{enumerate}
Each step can suffer from both experimental problems and theoretical
ambiguities.  Concentrating first on the theoretical ones, recently
these two steps have been fully realized for the chargino sector
\cite{choi1} and the work on exploiting the neutralinos is in progress
\cite{choi3}. Similar strategies have been  developed for sleptons and
squarks \cite{uli,Bartl}.  
An alternative approach for the step $B$, based only on
the masses of some of the charginos and neutralinos, can be found in
\cite{KM}.

\section{Determining the Lagrangian parameters}

\subsection{Charginos: extracting $\tan\beta$, $M_2$ and $\mu$}

The spin--1/2 superpartners of the $W$ boson and charged Higgs boson,
$\tilde{W}^\pm$ and $\tilde{H}^\pm$, mix to form chargino mass
eigenstates $\tilde{\chi}^\pm_{1,2}$. Their masses
$m_{\tilde{\chi}_{1,2}^\pm}$ and the mixing angles $\phi_L,\phi_R$ are
determined by the elements of the chargino mass matrix in the
$(\tilde{W}^-,\tilde{H}^-)$ basis \cite{mssm}
\begin{eqnarray}
{\cal M}_C=\left(\begin{array}{cc}
                M_2                &      \sqrt{2}m_W c_\beta  \\
             \sqrt{2}m_W s_\beta  &             \mu   
                  \end{array}\right)
\label{eq:mass matrix}
\end{eqnarray}
which is given in terms of fundamental parameters: $M_2$, $\mu$, and
$\tan\beta=v_2/v_1$; $s_\beta=\sin\beta$, $c_\beta=\cos\beta$. 
As outlined above, we will discuss first how to
determine the chargino masses and mixing angles [step $A$] and then
the procedure of extracting $M_2$, $\mu$, and $\tan\beta=v_2/v_1$
[step $B$].

Charginos are produced in $e^+e^-$ collisions, either in diagonal or
in mixed pairs
\begin{eqnarray*}
e^+e^- \ \rightarrow \ \tilde{\chi}^+_i \ \tilde{\chi}^-_j 
\end{eqnarray*}
With the second chargino $\tilde{\chi}_2^\pm$ expected to be
significantly heavier than the first one, at LEP2 or even in the first
phase of $e^+e^-$ linear colliders, the
chargino $\tilde{\chi}_1^\pm$ may be, for some time, the only chargino
state that can be studied experimentally in detail.  Therefore, we
concentrate on the diagonal pair production of the lightest chargino
$\tilde{\chi}_1^+\tilde{\chi}_1^-$ in $e^+e^-$ collisions.  Next,
assuming an upgrade in energy, we consider additional informations
available from $\tilde{\chi}^\pm_1 \ \tilde{\chi}^\mp_2$ and
$\tilde{\chi}^+_2 \ \tilde{\chi}^-_2$ production processes.

Two different matrices acting on the left-- and right--chiral
$(\tilde{W},\tilde{H})$ states are needed to diagonalize the
asymmetric mass matrix (\ref{eq:mass matrix}).  The two (positive)
eigenvalues are given by 
\begin{eqnarray}
m^2_{\tilde{\chi}^\pm_{1,2}}
  =\frac{1}{2}\left[M^2_2+\mu^2+2m^2_W
    \mp \Delta \, \right]
\end{eqnarray}
where
\begin{equation}
\Delta=\left[(M^2_2+\mu^2+2m^2_W)^2-4(M_2\mu-m^2_W\sin 2\beta)^2\right]^{1/2}
\end{equation} 
The left-- and right--chiral components of the mass eigenstate 
$\tilde{\chi}^-_1$ are related to the wino and higgsino components
in the following way,  
\begin{eqnarray}
&&\tilde{\chi}^-_{1L}=\tilde{W}^-_L\cos\phi_L
                     +\tilde{H}^-_{1L}\sin\phi_L \nonumber\\
&&\tilde{\chi}^-_{1R}=\tilde{W}^-_R\cos\phi_R
                     +\tilde{H}^-_{2R}\sin\phi_R 
\end{eqnarray}
with the rotation angles  given by
\begin{eqnarray}
&&\cos 2\phi_L=-(M_2^2-\mu^2-2m^2_W\cos 2\beta)/\Delta
\nonumber\\ 
&&\sin 2\phi_L=-2\sqrt{2}m_W(M_2\cos\beta+\mu\sin\beta)/\Delta
\nonumber\\
&&\cos 2\phi_R=-(M_2^2-\mu^2+2m^2_W\cos 2\beta)/\Delta
\nonumber\\ 
&&\sin 2\phi_R=-2\sqrt{2}m_W(M_2\sin\beta+\mu\cos\beta)/\Delta
\label{mixing}
\end{eqnarray}
As usual, we take $\tan\beta$ positive, $M_2$ positive and $\mu$ of either
sign.

\vspace{-5mm}
\begin{figure}[htb]
\hspace{1cm}
\epsfig{figure=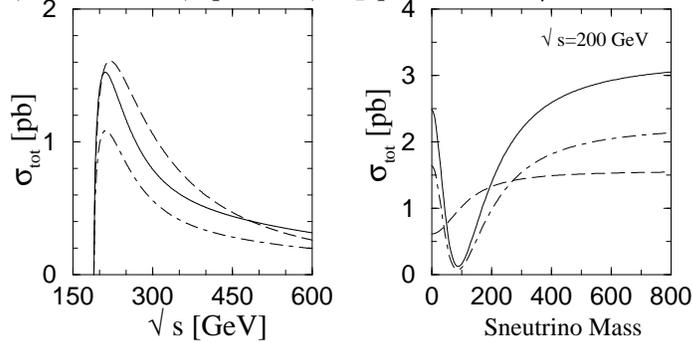, height= 2in}
\caption{Total cross section for the chargino pair production for a
  representative set of $M_2, \mu$: solid line for the gaugino case,
  dashed line for the higgsino case, dot-dashed line for the mixed
  case. In the left panel $m_{\tilde{\nu}}=200$ GeV (taken
  from~[14]).}
\label{fig:xrs1}
\end{figure}

Light charginos are produced in pairs in $e^+ e^-$ collisions through
s-channel $\gamma$ and $Z$, and t-channel sneutrino exchange.  The
production cross section will thus depend on the chargino mass
$m_{\tilde{\chi}^\pm_1}$, the sneutrino mass $m_{\tilde{\nu}}$ and the
mixing angles, eq.(\ref{mixing}), which determine the couplings of the
chargino states to the $Z$ and the sneutrino. The unpolarized total
cross section for $m_{\tilde{\chi}^\pm_1}=95$ GeV is illustrated in
fig.~\ref{fig:xrs1} for representative cases of dominant higgsino,
gaugino or mixed content of the lightest chargino state. The sharp
rise near threshold should allow a precise determination of the
chargino mass.  The sensitivity to the sneutrino mass with the typical
destructive interference in the gaugino and mixed cases necessitates
the knowledge of this parameter \cite{gudi}.

Charginos are not stable and each will decay directly to a pair of
matter fermions (leptons or quarks) and the (stable) lightest
neutralino $\tilde{\chi}_1^0$.  The decay proceeds through the
exchange of a $W$ boson (charged Higgs exchange is suppressed for
light fermions) or scalar partners of leptons or quarks. The decay
matrix elements will depend on further parameters like the scalar
masses and couplings to the neutralino. In addition, the presence of
two invisible neutralinos in the final state of the process, $ e^+ e^-
\to \tilde{\chi}_1^+ \tilde{\chi}_1^- \to \tilde{\chi}_1^0
\tilde{\chi}_1^0 (f_1 \bar{f}_2) (\bar{f}_3 f_4)$, makes it impossible
to measure directly the chargino production angle $\Theta$ in the
laboratory frame.  Integrating over this angle and also over the
invariant masses of the fermionic systems $(f_1 \bar{f}_2)$ and
$(\bar{f}_3 f_4)$, one can write the differential cross section in the
following form:
\begin{equation}
\frac{d^4 \sigma( e^+ e^- \to \tilde{\chi}_1^0 
\tilde{\chi}_1^0 (f_1 \bar{f}_2)
(\bar{f}_3 f_4)) }{d\cos \theta^* d\phi^* d\cos \bar{\theta}^*
d\bar{\phi}^*} = \frac{\alpha^2 \beta}{124 \pi s} {\cal B}\, \bar{\cal B}
\, \Sigma(\theta^*, \phi^*, \bar{\theta}^*, \bar{\phi}^*)
\end{equation}
where $\alpha$ is the fine structure constant, $\beta$ the velocity of
the chargino in the c.m. frame. For the $\tilde{\chi}^-_1$ decay we
have ${\cal B}=Br(\tilde{\chi}_1^- \to \tilde{\chi}_1^0 f_1
\bar{f}_2)$, $\theta^*$ is the polar angle of the $f_1 \bar{f}_2$
system in the $\tilde{\chi}^-_1$ rest frame with respect to the
chargino's flight direction in the lab frame, and $\phi^*$ is the
azimuthal angle with respect to the production plane; quantities with
a bar refer to the $\tilde{\chi}^+_1$ decay.  The differential cross
section $\Sigma(\theta^*, \phi^*, \bar{\theta}^*, \bar{\phi}^*)$ is
expressed in terms of sixteen independent angular combinations of
helicity production amplitudes
\begin{equation}
  \begin{array}{rcl} \Sigma & = & \Sigma_{unpol} + \kappa \cos
\theta^* {\cal P} + \bar{\kappa}\cos \bar{\theta}^* \bar{{\cal P}} +
\cos \theta^* \cos \bar{\theta^*} \kappa \bar{\kappa} {\cal Q} \\[2mm]
&& + \sin \theta^* \sin \bar{\theta^*} \cos (\phi^* + \bar{\phi^*})
\kappa \bar{\kappa} {\cal Y} + \dots
\end{array} \label{Sigma}
\end{equation}
Out of the sixteen terms, corresponding to the unpolarized,
$2 \times 3$ polarization components and $3 \times 3$ spin--spin
correlations in the production process, only 7 are independent 
(neglecting small effects from the $Z$-boson width
and loop corrections) and $\kappa=-\bar{\kappa}$ in the
CP-invariant theory.
The polarization component ${\cal P}$ coming  
from the $\tilde{\chi}_1^-$ system, for example,  reads
\begin{equation}
{\cal P}={\textstyle \frac{1}{4}}\int{\rm d}\cos\Theta\sum_{\sigma=\pm}
      [|A_{\sigma;++}|^2+|A_{\sigma;+-}|^2
           -|A_{\sigma;-+}|^2-|A_{\sigma;--}|^2 ]
\end{equation}
where $2\pi\alpha A_{\sigma;\lambda\lambda'}$ is the helicity
amplitude with $\sigma;\lambda\lambda'$ denoting the helicities of the
electron and $\tilde{\chi}^-_1\tilde{\chi}^+_1$ pair, respectively.
All the complicted dependence on the chargino decay dynamics
(neutralino and sfermion masses and their couplings) is contained in
the spin analysis-powers $\kappa$ and $\bar{\kappa}$.

\vspace{2cm}
\begin{figure}[htb]
\centerline{ \epsfig{figure=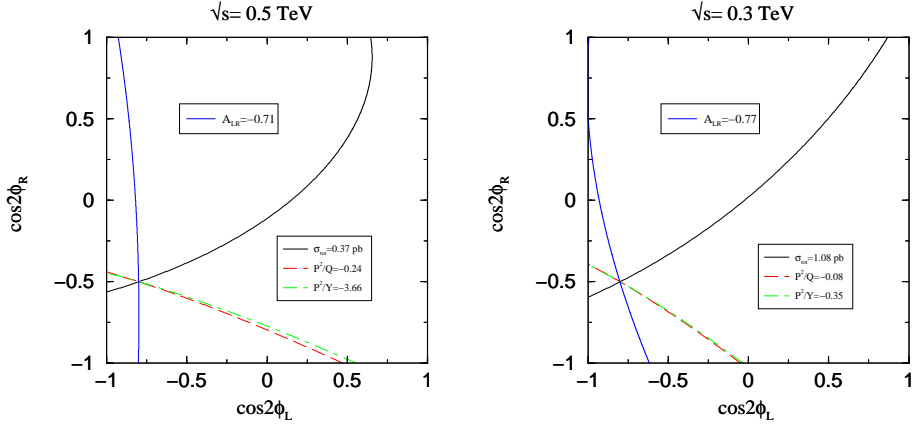, height=4cm}}
\caption{Contours for the ``measured values'' of the total cross
  section (solid line), ${\cal P}^2/{\cal Q}$, and ${\cal P}^2/{\cal
  Y}$ (dot-dashed line) for $m_{\tilde{\chi}_1^{\pm}}=95 GeV$
  [$m_{\tilde{\nu}} = 250$ GeV]. Superimposed are contour lines
  (solid, almost vertical lines) for the ``measured'' LR asymmetry. }
\label{fig:cont}
\end{figure}

The crucial observation of \cite{choi1} is that all explicitly written
terms in eq.~(\ref{Sigma}) can be extracted and three
$\kappa$-independent physical observables, $\Sigma_{unpol}, {\cal
P}^2/{\cal Q}$ and ${\cal P}^2/{\cal Y}$, constructed. Indeed, it is
possible by means of kinematical projections, since $\cos \theta^*, \cos
\bar{\theta^*}$ and $\sin \theta^* \sin \bar{\theta^*} \cos( \phi^* +
\bar{\phi^*})$ are fully determined by the measurable parameters $E,
|\vec{p}|$ (the energy and momentum of each of the decay systems $f_i
\bar{f_j}$ in the laboratory frame) and the chargino mass.  As a
result, the chargino properties can be determined {\it independently}
of the other sectors of the model.\footnote{Actually to determine the
kinematical variables, $\cos \theta^*$ etc., the knowledge of
$m_{\tilde{\chi}^0_1}$ is also needed, which can be extracted from the
energy distributions of final state particles, see later.  However, it
must be stressed that the above procedure does not depend on the
details of decay dynamics nor on the structure of (potentially more
complex) neutralino and sfermion sectors.}  The measurements of the
cross section and either of the ratios ${\cal P}^2/{\cal Q}$ or ${\cal
P}^2/{\cal Y}$ can be interpreted as contour lines in the plane
$\{\cos 2\phi_L,\cos 2\phi_R\}$ which intersect with large angles so
that a high precision in the resolution can be achieved. A
representative example for the determination of $\cos 2\phi_L$ and
$\cos 2\phi_R$ is shown in fig.~\ref{fig:cont}. The mass of the light
chargino is set to $m_{\tilde{\chi}^\pm_1}=95$ GeV, and the
``measured'' cross section, ${\cal P}^2/{\cal Q}$ and ${\cal
P}^2/{\cal Y}$ at $\sqrt{s}=500$ GeV are taken to be
\begin{equation}
\sigma(e^+e^-\rightarrow\tilde{\chi}^+_1\tilde{\chi}^-_1)=0.37\ \ 
{\rm pb},\quad
{\cal P}^2/{\cal Q}=-0.24,\quad 
{\cal P}^2/{\cal Y}= -3.66
\label{eq:measured} 
\end{equation}
in the left panel of fig.~\ref{fig:cont}.  The three contour lines
meet at a single point $\{\cos 2\phi_L,\cos 2\phi_R\}
=\{-0.8,-0.5\}$. The sneutrino mass is set to $m_{\tilde{\nu}}=250$
GeV.  Note that the $m_{\tilde{\nu}}$ can be determined together with
the mixing angles by requiring a consistent solution from the
``measured quantities'' $\sigma$, ${\cal P}^2/{\cal Q}$ and ${\cal
P}^2/{\cal Y}$ at several values of incoming energy, as exemplified in
both panels of fig.~\ref{fig:cont} for $\sqrt{s}=500$ and 300 GeV.

If polarized beams are available, the left-right asymmetry $A_{LR}$
can provide an alternative way to extract the mixing angles (or serve 
as a consistency check). This is also demonstrated in
fig.~\ref{fig:cont}, where contour lines for the ``measured'' values of
$A_{LR}$ are also shown.  Moreover, with right-handed electron beams
one can turn off the sneutrino exchange in the production process and
since at high energy the $\gamma$ and $Z$ ``demix'' back to the
$W^0_3$ and $B^0$ gauge bosons, only the higgsino component of the
chargino is selected. Thus the polarization alone will give us the
composition of charginos.  In short, the step $A$ can be fully
realized for the lightest charginos.

Let us now discuss the step $B$ and  describe briefly how to determine the
Lagrangian parameters $M_2, \mu$ and $\tan\beta$ from 
$m_{\tilde{\chi}^\pm_1}$, 
$\cos 2\phi_L$ and $\cos 2\phi_R$. It is most
transparently achieved by introducing the two triangular quantities
\begin{eqnarray}
p=\cot(\phi_R-\phi_L)\ \ {\rm and}\  \ q=\cot(\phi_R+\phi_L)
\end{eqnarray}
They are expressed in terms of the measured values $\cos 2\phi_L$ and
$\cos 2\phi_R$  up to a discrete ambiguity due to undetermined signs 
 of $\sin 2\phi_L$ and $\sin 2\phi_R$
\begin{eqnarray}
p^2+q^2 &=&\frac{2(\sin^2 2\phi_L+\sin^2 2\phi_R)}{(\cos 2\phi_L-\cos 2
\phi_R)^2}
           \nonumber\\
  pq    &=&\frac{\cos 2\phi_L+\cos 2\phi_R}{\cos 2\phi_L-\cos 2\phi_R}
\nonumber\\
p^2-q^2 &=&\frac{4\sin 2\phi_L\sin 2\phi_R}{(\cos 2\phi_L-\cos 2\phi_R)^2} 
\end{eqnarray}
Solving then eqs.~(\ref{mixing}) for $\tan\beta$ one finds at most 
two possible solutions, and using 
\begin{eqnarray}
M_2&=&{m_W}[(p+q)s_\beta-(p-q)c_\beta]/\sqrt{2}
       \nonumber\\
\mu&=&{m_W}[(p-q)s_\beta-(p+q)c_\beta]/\sqrt{2}
\label{eq:M2_mu}
\end{eqnarray}
we arrive at $\tan\beta$, $M_2$ and $\mu$ up to a two-fold
ambiguity. For example, taking the ``measured values'' from
eq.~(\ref{eq:measured}), the following results are found in
\cite{choi1}
\begin{eqnarray}
 [\tan\beta; M_2,\mu] =
        \left\{\begin{array}{l}
              [1.06; \;  83{\rm GeV}, \; -59{\rm GeV}] \\
                  { }\\ {}
              [3.33; \;  248{\rm GeV}, \; 123{\rm GeV}]
              \end{array}\right.
\end{eqnarray}
To summarize, from the lightest chargino pair production, the 
measurements of the total production cross section and either the
angular correlations among the chargino decay products (${\cal
P}^2/{\cal Q}$, ${\cal P}^2/{\cal Y}$) or the LR asymmetry, the step
$A$ can be realized and the physical parameters
$m_{\tilde{\chi}^\pm_1}$, $\cos 2\phi_L$ and $\cos 2\phi_R$ determined
unambiguously. Then the fundamental parameters $\tan\beta$,
$M_2$ and $\mu$ are extracted (step $B$) up to a two-fold ambiguity.

If the collider energy is sufficient to produce the two chargino
states in pairs, the above ambiguity can be removed \cite{choi2}. 
The new ingredient in this case is the knowledge of the
heavier chargino mass. Like for the ligter one,
$m_{\tilde{\chi}^\pm_{2}}$ can be determined very precisely from the
sharp rise of the production cross sections
$\sigma(e^+e^-\rightarrow\tilde{\chi}^-_i\tilde{\chi}^+_j)$.
Then the value of $\tan\beta$ is uniquely determined in terms of the
mass difference of two chargino states, $\Delta =
m^2_{\tilde{\chi}^\pm_2}-m^2_{\tilde{\chi}^\pm_1}$, and two mixing
angles as follows
\begin{eqnarray}
\tan\beta=\sqrt{\frac{4m^2_W
                     +\Delta \,
                      (\cos 2\phi_R-\cos 2\phi_L)}{4m^2_W
                     -\Delta \,
                      (\cos 2\phi_R-\cos 2\phi_L)}}
\label{eq:tanb}
\end{eqnarray}
Using the convention $M_2>0$, 
    the gaugino mass parameter $M_2$ and the modulus of the higgsino
    mass parameter are given by
\begin{eqnarray}
 M_2&=&\frac{1}{2}
        \sqrt{2(m^2_{\tilde{\chi}^\pm_2}+m^2_{\tilde{\chi}^\pm_1}-2m^2_W)
              -\Delta \, 
               (\cos 2\phi_R+\cos 2\phi_L)}\nonumber\\
 |\mu|&=&\frac{1}{2}
        \sqrt{2(m^2_{\tilde{\chi}^\pm_2}+m^2_{\tilde{\chi}^\pm_1}-2m^2_W)
              +\Delta \, 
               (\cos 2\phi_R+\cos 2\phi_L)}
\label{eq:M2mu}
\end{eqnarray} 
The sign of $\mu$ is then determined by the sign of the following 
expression
\begin{eqnarray}
\mbox{sign}(\mu)= \mbox{sign}[ \Delta^2
                   -(M^2_2-\mu^2)^2-4m^2_W(M^2_2+\mu^2)
                   -4m^4_W\cos^2 2\beta]
\end{eqnarray}

Before leaving the chargino sector, let us note that from the energy
distribution of the final particles in the decay of the charginos
$\tilde{\chi}^\pm_1$, the mass of the lightest neutralino
$\tilde{\chi}^0_1$ can be measured \cite{uli}. 
This, as we will see in the next subsection, 
allows us to derive the parameter $M_1$ 
in the CP--invariant theories so that 
the neutralino mass matrix, too, can be reconstructed
in a model-independent way. 

\subsection{And Neutralinos: extracting also $M_1$}

The spin--1/2 superpartners of the neutral electroweak gauge bosons
and neutral Higgs bosons mix to form four neutralino mass eigenstates
$\tilde{\chi}^0_{1,2,3,4}$. Their masses $m_{\tilde{\chi}_{i}^0}$ and
the mixing angles are determined by the elements of the neutralino
mass matrix given by ($s_W=\sin\theta_W$, $c_W=\cos\theta_W$) \cite{mssm}
\begin{eqnarray}
{\cal M}_N = \left[ \begin{array}{cccc}
M_1 & 0 & -m_Z s_W c_\beta & m_Z  s_W s_\beta \\
0   & M_2 & m_Z c_W c_\beta & -m_Z  c_W s_\beta \\
-m_Z s_W c_\beta & m_Z  c_W c_\beta & 0 & -\mu \\
m_Z s_W s_\beta & -m_Z  c_W s_\beta & -\mu & 0
\end{array} \right]
\end{eqnarray}
Since ${\cal M}_N$ is symmetric, 
an orthogonal matrix ${\cal N}$  can be constructed that transforms 
${\cal M}_N$ to a (positive) diagonal matrix. This 
mathematical problem can be solved analytically \cite{esa}.
Due to the large ensemble of four neutralinos, however, the analysis is much
more complex than in the chargino case.  In particular, the step $B$,
$i.e.$ the analytical reconstruction of the fundamental SUSY
parameters, is more complicated although,
after measuring the parameters $M_2, \mu$ and $\tan \beta$ from the
chargino production, the only additional parameter in the neutralino
mass matrix is $M_1$.

Neutralinos are produced in $e^+e^-$ collisions either in diagonal or
non-diagonal pairs. The lightest neutralino $\tilde{\chi}^0_1$ is
generally expected to be the lightest SUSY particle (LSP) and
therefore stable in the $R$-parity preserving model. As a result, the
production of the lightest neutralino pairs is difficult to identify
and exploit experimentally.  Therefore we consider production
processes where at least one of the neutralinos is not an LSP, for
example $\tilde{\chi}^0_1\tilde{\chi}^0_2$ or
$\tilde{\chi}^0_2\tilde{\chi}^0_2$. These processes are generated by
the $s$-channel $Z$ exchange and the $t$- and $u$-channel selectron
$\tilde{e}_{L,R}$ exchanges. The transition matrix elements will then
depend not only on the neutralino properties but on the selectron
masses as well.  The heavier neutralino $\tilde{\chi}^0_2$ will decay
into the LSP and a fermion pair, leptons or quarks,
$\tilde{\chi}^0_2\rightarrow \tilde{\chi}^0_1 f\bar{f}$, throught the
exchange of a $Z$ boson or scalar partners of the fermion (the neutral
Higgs boson exchange is negligible for light fermions).  The decay
products will serve as a signature of the production process and from
the fast rise of the cross sections the masses $m_{\tilde{\chi}^0_i}$
can be measured precisely.

Additional informations can be obtained by analysing the angular
correlations among the decay products, like in the chargino sector. In
the case of $\tilde{\chi}^0_2\tilde{\chi}^0_2$, the method developed
for the chargino case can be applied directly. One can attempt to
separate the production from the decay processes and determine the
$Z\tilde{\chi}_i^0\tilde{\chi}_j^0$ and $e\ti e \tilde{\chi}_i^0$ couplings
(expressed as known combinations of the mixing matrix elements ${\cal
N}_{ij}$ \cite{mssm}). Such a separation is interesting for the hadronic
or $\mu^+\mu^-$ decay modes of $\tilde{\chi}^0_2$ ($\tilde{\chi}^0_2
\rightarrow \tilde{\chi}^0_1 q\bar{q}$ or $\tilde{\chi}^0_1 \mu^+\mu^- $) 
because independent
information on the neutralino couplings to electron-selectron and
quark-squark or $\mu^\pm\tilde{\mu}^\mp$ from the production and decay 
processes, respectively,
can be inferred.  For the $\tilde{\chi}^0_2
\rightarrow \tilde{\chi}^0_1 e^+e^-$ the production/decay separation 
might be useful only from the point of view of consistency
checks. In the $\tilde{\chi}^0_1\tilde{\chi}^0_2$ production case, only
one neutralino decays and such a separation is not possible due to a
limited number of measurable kinematical variables. 
Nevertheless some information on couplings can be extracted. The 
prescription  for the step $A$ in the neutralino sector  still awaits 
a detailed analysis \cite{choi3}.

\begin{figure}
\center
\epsfig{figure=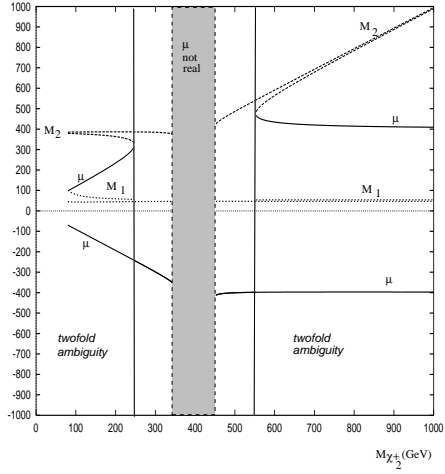, height= 2.5in}
\caption{$\mu, M_1$ and $M_2$ (with the ``higgsino-like'' convention
  $|\mu| \leq M_2$) as functions of $m_{\tilde{\chi}^+_2}$ for fixed
  $m_{\tilde{\chi}^+_1} = 400$ GeV, $m_{\tilde{\chi}^0_2} = 50$ GeV,
  and $\tan\beta = 2$ (taken from~[18]). }
\label{fig:km}
\end{figure}

However, the knowledge of $m_{\tilde{\chi}^0_1}$ in addition to the
measurements performed in the chargino sector is sufficient to
pinpoint the value of $M_1$, the only new parameter. This particular
problem has recently been considered in \cite{KM}, where the emphasis
has been put on the step $B$, namely to what extent
the reconstruction of the 
Lagrangian parameters through
a controllable analytical procedure,  including all possible ambiguities, 
is possible if three of the chargino and neutralino 
masses and $\tan \beta$ were known. Two cases have been analysed:
\begin{itemize}
\item[$S_1$:] the two charginos and one neutralino masses are input,
\item[$S_2$:] one chargino and two neutralino masses are input.
\end{itemize}
In case $S_1$, a closed analytical procedure to determine $M_1$
has been found. The crucial observation is to use the four independent
linear combinations of the entries of ${\cal M}_N$ which are invariant
under similarity transformations, and thus relate them simply to the
four eigenvalues of ${\cal M}_N$. As a result, any of the neutralino
masses taken as input, for example $\tilde{\chi}^0_1$, in addition to
$\mu$, $M_2$ and $\tan\beta$ allows the set of these 
consistency relations to be
solved for the other three neutralino masses. Then the $M_1$ parameter
is determined as
\begin{equation}
 M_1= 
   -\frac{P_{i j}^2 + 
       P_{i j} (\mu^2 + m_Z^2 + M_2 S_{i j} - S_{i j}^2) 
              + \mu m_Z^2 M_2 s_W^2 \sin 2 \beta}
      {P_{i j} (S_{i j} -M_2) + \mu ( c_W^2 m_Z^2 \sin 2 \beta -\mu M_2 )}
\end{equation}
where
\begin{eqnarray} 
S_{i j} \equiv \tilde{m}_{i} + \tilde{m}_{j}, \qquad 
P_{i j} \equiv \tilde{m}_{i}  \tilde{m}_{j} \nonumber 
\end{eqnarray}
$i\neq j$, and $\tilde{m}_i=\epsilon_i m_{\tilde{\chi}^0_i}$ (the mass
parameters can be negative, for the details we refer to \cite{KM}).
As an example of the numerical result of such a procedure, the sensitivity to a chargino mass with the other chargino and the neutralino masses fixed
is shown in fig.~\ref{fig:km}. 

In  case  $S_2$, the above consistency relations can be
reformulated in terms of two quadratic equatiuons for $M_2$ and $M_1$
at a given value of $\mu$ (and $\tan\beta$). Without any additional
theoretical input, a numerical (iterative) procedure is used to obtain
at most four distinct solutions for $\mu$, $M_1$ and $M_2$ for a given
set of ${\tilde{\chi}^\pm_1}$ and two neutralino masses.

\subsection{Sfermions: extracting $m_{\tilde{f}_L}$, $m_{\tilde{f}_R}$ 
and $A^f$}

For each fermion chirality $f_{L,R}$ supersymmetry predicts a
corresponding sfermion $\tilde{f}_{L,R}$.  Since SUSY is broken, the
chiral left and right sfermions $\tilde{f}_L$ and $\tilde{f}_R$ may
aquire different mass terms and they can mix.  The mass eigenstates
and mixing are determined by the mass matrices (for a given sfermion
flavor $\tilde{f}$)
\begin{eqnarray}
M^2_{\tilde{f}} = 
\left[ \begin{array}{cc} m_{\tilde{f}_L}^2 + m_f^2 & m_f (A^f - \mu r_f) 
\\ m_f (A^f - \mu r_f)  & m_{\tilde{f}_R}^2 + m_f^2 \end{array} \right]
\label{sfmass}
\end{eqnarray}
with
\begin{eqnarray}
  &&\hspace{-8mm} m^2_{\ti f_L} =
  m^2_{\ti Q} + m_Z^2 \cos 2\beta \,(T^3_f - e_f \sin^2\theta_W) ,
  \label{eq:msfl}\\
  &&\hspace{-8mm} m^2_{\ti f_R} =
  m^2_{\ti F'} + e_f m_Z^2 \cos 2\beta \sin^2\theta_W ,
  \label{eq:msfr}
\end{eqnarray}
where $e_f$ and $T^3_f$ are the charge and the third component
of the weak isospin of the sfermion $\ti f$,
 $m_f$ is the mass of the corresponding fermion,  
$m_{\ti F'} = m_{\ti u}$, $m_{\ti u}$ for $\ti f_R = \ti u_R$, 
$ \ti d_R$, respectively, and 
$r_{f} = 1/\tan\beta$  for up- and $r_{f} = \tan\beta$  for 
down-type sfermions. 
The matrices (\ref{sfmass}) are diagonalized by orthogonal 
transformations with mixing angles
$\theta_f$ defined by
\begin{eqnarray}
\sin 2\theta_f = \frac{2 m_f (A^f -\mu r_f)} { m_{\tilde{f}_1}^2
-m_{\tilde{f}_2}^2 } \ \ , \ \ 
\cos 2\theta_f = \frac{m_{\tilde{f}_L}^2 -m_{\tilde{f}_R}^2}
{ m_{\tilde{f}_1}^2 -m_{\tilde{f}_2}^2 } 
\end{eqnarray}
and the masses of the sfermion eigenstates are given by
\begin{eqnarray}
m_{\tilde{f}_{1,2}}^2 = m_f^2 + \frac{1}{2} \left[ 
m_{ \tilde{f}_L}^2 + m_{\tilde{f}_R}^2 \mp \sqrt{
(m_{\tilde{f}_L}^2 - m_{\tilde{f}_R}^2)^2 + 4m_f^2 (A^f -\mu r_f)^2 } 
\right].
\end{eqnarray}
Since the mixing term is of order $m_f$, it can be substantial 
only for the third generation sfermions (for sbottom and stau if
$\tan\beta$ is large), with an important consequence of lowering the
mass of the lighter eigenstate $m_{\tilde{f}_{1}}$.  As a result, the
lighter stop ${\tilde{t}_{1}}$ is expected to be the lightest scalar
fermion.

Sfermions are produced in pairs in $e^+e^-$ collisions
\begin{equation}
e^+e^- \ra \ti f_{i} \bar{\ti f}_j
\end{equation} 
through $s$-channel $\gamma$ and $Z$ exchange; only the selectron
production receives an additional $t$-channel neutralino exchange
contribution. Since the gauge boson couplings respect chirality, the
nondiagonal $\ti f_{1} \bar{\ti f}_2$ production can occur only 
for nontrivial mixing.

It is important first to verify experimentally the chiral nature of
produced sfermions. This can be easily done at $e^+e^-$ colliders by
using polarized beams (not available at LHC) or by reconstructing the
polarization of the final state fermions from sfermion decays (too
difficult in hadron collisions).  As an example, consider the pair
production of right-handed staus which most probably will decay into
the LSP neutralinos and $\tau$'s. The signature is the same as that of
$W$ pair production with the $W$'s decaying into $\tau\nu_\tau$.
However, at high-energy the $Z$ and $\gamma$ ``demix'' back to the
$W_3^0$ and $B^0$ (hypercharge). Since the former does not couple to
right-handed states, only the hypercharge boson is exchanged in right
sfermion production.  Therefore the background $W$ pair production can
be suppressed by choosing right-handed electrons.  Moreover, as a
result of hypercharge assignements $Y(e_L)=-1$ and $Y(e_R)=-2$, the
signal cross section with right-handed $e^-$ beams will be by factor 4
larger than with the left-handed $e^-$ beams.  The beam polarization
therefore is a very powerful tool: allows us not only to tag the
nature of the stau (right-handed) independently of its decays and
increase the signal cross section, but also suppress the
background. All these has been checked by the full simulation of the
Japanese group \cite{JapanSusy}.  In addition, reconstruction of the
$\tau$ polarization in the decay process $\stau\ra \tilde{\chi}^0_1
\tau$ will play an important role in exploring the Yukawa couplings.
The $\stau\tau\tilde{\chi}^0_1$ coupling depends on the neutralino
composition. The interaction involving gaugino component ($\tilde{B}$
or $\tilde{W}$) is proportional to gauge couplings and is chiral
conserving, whereas the interaction involving higgsino component
($\tilde{H}_{1,2}$) is proportional to $\tau$ Yukawa coupling
$Y_{\tau}\sim m_{\tau}/\cos\beta$ and chiral flipping. Thus the
polarization of $\tau$ lepton from $\ti \tau$ decays depends on the
ratio of the chirality flipping and chirality concerving interactions,
and consequently on $\tan\beta$. For a detailed discussion of stau
production we refer to \cite{Nojiri}.

Once the sfermion production has been optimized, one can either infer
the sfermion mass from a threshold scan (which is independent of the
decay) or (as in chargino case) the measurement of the fermion energy
spectrum will give both the $m_{\ti f}$ and the LSP mass. A combined
fit for a low luminosity option of 10 fb$^{-1}$ and 85\% polarization
of the electron beam shows that a precision of order a few percent for
sfermion masses can easily be obtained \cite{uli}.

\begin{figure}
\center
\epsfig{figure=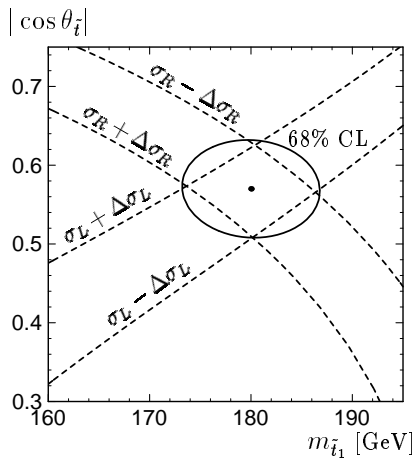, height= 2.5in}
\caption{Error bands (dashed) and the corresponding error ellipse as a
  function of $m_{\st_1}$ and $|\cos\theta_{\st}|$ for the tree--level
  cross sections of $e^+ e^- \to \st_1 \st_1$ at $E_{cm} =
  500$~GeV with 90\% left-- and right--polarized electron beam (taken
  from~[17]). }
\label{sab}
\end{figure}

A case study of $e^+ e^- \to \st_1 \bar\st_1$ 
with the aim of determining the SUSY parameters has been performed by
the Vienna group \cite{Bartl} at $\sqrt s =$~500~GeV and 
${\cal L}=50$ fb$^{-1}$. 
The input $m_{\st_1} =$~180~GeV and  
left--right stop mixing angle $|\cos\tht|$ = 0.57  
corresponds to
the minimum of the cross section. The cross sections at tree level for
these parameters are $\sigma_L =$ 48.6~f\/b and $\sigma_R =$ 46.1~f\/b
for 90\% left-- and right--polarized $e^-$~beam,
respectively.  Based on detailed studies 
the experimental errors on these cross sections are
estimated to be  $\Delta \sigma_L =
\pm$~6~f\/b and $\Delta \sigma_R = \pm$~4.9~f\/b.  Figure~\ref{sab} shows the
resulting error bands and the corresponding error ellipse in the
$m_{\st_1}$--$\cos\tht$ plane.  The experimental accuracy for the stop
mass and mixing angle are $m_{\st_1} = 180 \pm 7$~GeV, $|\cos\tht| =
0.57 \pm 0.06$.

Additional experimental input is needed, however, to determine the
fundamental parameters. The Vienna group decided to exploit the
sbottom system. Assuming that $\tan\beta$ is low and the
$\sb_L$--$\sb_R$ mixing can be neglected, $i.e.$ $\cos\theta_\sb = 1$, 
and taking $\sb_1 = \sb_L=200$ GeV, $\sb_2 =
\sb_R=220$ GeV,  the cross sections and the expected
experimental errors are $\sigma_L(e^+ e^- \to \sb_1 \bar\sb_1) = 61.1
\pm 6.4$~f\/b, $\sigma_R(e^+ e^- \to \sb_2 \bar\sb_2) = 6 \pm
2.6$~f\/b for the 90\% left-- and right--polarized $e^-$ beams.
The resulting experimental errors are $m_{\sb_1} = 200 \pm 4$~GeV,
$m_{\sb_2} = 220 \pm 10$~GeV. 
With these results
the mass of the
heavier stop can be calculated and is found to be 
$m_{\st_2} = 289 \pm 15$~GeV. 
Verifying this prediction experimentally  will test the MSSM.  

To complete the step $B$,  $\mu = - 200$~GeV, $\tan\beta = 2$ and  
$m_t = 175$~GeV
have been taken assuming that $\mu$ and $\tan\beta$ are known 
from other experiments (from chargino sector, for example).
The soft-supersymmetry breaking parameters of the stop and sbottom systems 
can then be determined up to a two-fold ambiguity:
$m_{\tilde Q} = 195 \pm 4$~GeV, $m_{\tilde u} = 138 \pm 26$~GeV,
$m_{\tilde d} = 219 \pm 10$~GeV, $A^t = -236 \pm 38$~GeV 
if $\cos\tht > 0$,
and $A^t = 36 \pm 38$~GeV if $\cos\tht < 0$.

\section{Conclusions}
In this talk I have tried to illustrate the discovery power and
precision tools developed to explore supersymmetry at future $e^+e^-$
linear colliders. The LC is an excellent machine for supersymmetry
because a systematic, model-independent determination of the
supersymmetry parameters is possible within a discovery reach that is
limited by the available center-of-mass energy.  Although we only
considered real-valued parameters, some of the material presented here
goes through unaltered if phases are allowed \cite{choi1,choi2} even
though extra information will still be needed to determine those
phases.

It should be stressed that the strategies presented here are just at
the theoretical level. A more realistic simulation of the experimental
measurements of physical observables and related errors is still
needed to assess fully the physics potential of LC. 
Nevertheless, if
the LC and detectors are built and work as expected, I have no doubt
that the actual measurements will be better than anything I have
presented here -- provided supersymmetry is discovered!  After all,
nobody beats experimentalists with real data.

\section*{Acknowledgements}
I would like to thank Gudrid Moortgat-Pick, Stefan Pokorski and Peter
Zerwas for many valuable discussions. This work has been partially
supported by the KBN grant 2 P03B 052 16.

\end{document}